\documentclass[sigconf]{acmart}

\AtBeginDocument{%
  }

\setcopyright{acmlicensed}
\copyrightyear{2018}
\acmYear{2018}
\acmDOI{XXXXXXX.XXXXXXX}
\acmConference[Conference acronym 'XX]{Make sure to enter the correct
  conference title from your rights confirmation email}{June 03--05,
  2018}{Woodstock, NY}

\acmISBN{978-1-4503-XXXX-X/2018/06}

\usepackage[table]{xcolor}
\usepackage{amsmath,amsfonts}
\usepackage{graphicx}
\usepackage{textcomp}
\usepackage{algorithm}
\usepackage{algpseudocode}
\usepackage{algorithmicx}
\usepackage{diagbox}
\usepackage{booktabs}
\usepackage{tcolorbox}
\usepackage{tabularx}
\usepackage{soul}
\usepackage{balance}
\usepackage{array}
\usepackage[table]{xcolor}
\definecolor{bestgreen}{HTML}{DDE9B2}
\definecolor{secondblue}{HTML}{DCE6F1}
\definecolor{oraclered}{HTML}{FDC1C1}

\usepackage{amssymb}
\newcommand{\green}[1]{\colorbox{bestgreen}{\strut #1}}
\newcommand{\blue}[1]{\colorbox{secondblue}{\strut #1}}

\newcommand{\best}[1]{\cellcolor{bestgreen}\textbf{#1}}
\newcommand{\second}[1]{\cellcolor{secondblue}\textbf{#1}}
\newcommand{\oracle}[1]{\cellcolor{oraclered}{\strut #1}}
\usepackage{algorithm}      
\usepackage{algpseudocode}  
\usepackage{algorithmicx}   
\usepackage{framed}         
\usepackage{mdframed}       

\usepackage{framed}
\usepackage{amsmath}

\definecolor{mysoftred}{HTML}{FFCCCC}
\newcommand{\hlstate}[1]{\colorbox{mysoftred}{\strut #1}}

\usepackage{makecell}
\usepackage{graphicx}
\usepackage{subfig}
\usepackage{blindtext}
\usepackage{adjustbox}
\usepackage{multirow}
\usepackage{color}
\usepackage{booktabs}
\usepackage{tabularx}
\usepackage{colortbl}

\usepackage{bbding}

\usepackage{tikz}
\usepackage{listings}
\usepackage{etoolbox}
\usepackage{subfig}
\usepackage{url}
\usepackage{setspace}
\usepackage[english]{babel}
\usepackage{balance}
\usepackage{array}

\usepackage{xcolor}

\definecolor{Gray}{gray}{0.9}

\newcommand{\circled}[2][]{\tikz[baseline=(char.base)]
	{\node[shape = circle, draw, inner sep = 1pt]
		(char) {\phantom{\ifblank{#1}{#2}{#1}}};%
		\node at (char.center) {\makebox[0pt][c]{#2}};}}
\robustify{\circled}

\usepackage{xcolor}
\definecolor{fixmeyellow}{HTML}{FFF59D}

\newcommand{\SystemName}{\textsc{AutoPass}\xspace}

\newcommand\cparagraph[1]{\vspace{1.5mm} \noindent \textbf{#1}\xspace}

\author{Zepeng Li}
\affiliation{%
  \institution{Shaanxi Normal University}
  \country{China}
}

\author{Jie Ren}
\affiliation{%
  \institution{Shaanxi Normal University}
  \country{China}
}
\email{renjie@snnu.edu.cn}

\author{Zhanyong Tang}
\affiliation{%
  \institution{Northwest University}
  \country{China}
}

\author{Jie Zheng}
\affiliation{%
  \institution{Northwest University}
  \country{China}
}

\author{Zheng Wang}
\affiliation{%
  \institution{University of Leeds}
  \country{United Kingdom}
}

\begin{document}

\title{\SystemName: Evidence-Guided LLM Agents for Compiler Performance Tuning}

\renewcommand{\shortauthors}{Trovato et al.}

\begin{abstract}
Large Language Models (LLMs) show promise for code compilation tasks, but applying them to runtime performance tuning is difficult due to complex microarchitectural effects and noisy runtime measurements. We present \SystemName, a multi-agent framework for compiler performance tuning that uses compiler and runtime evidence to guide LLM-generated optimization decisions. Rather than treating the compiler as a black box like prior auto-tuning schemes, \SystemName opens up the compiler to the LLM, enabling it to query compiler-internal optimization states and analyze the intermediate representation to orchestrate compiler options. The search process iteratively refines optimization configurations using measured runtime feedback to diagnose regressions and guide latency-improving edits. \SystemName operates in an inference-only, training-free setting and requires no offline training or task-specific fine-tuning, making it readily applicable to new benchmarks and platforms.
We implement \SystemName on the LLVM compiler and evaluate it on server-grade x86-64 and embedded ARM64 systems. \SystemName outperforms expert-tuned heuristics and classical autotuning methods, achieving geometric-mean speedups of 1.043$\times$ and 1.117$\times$ over LLVM \texttt{-O3} on x86-64 and ARM64, respectively.

\end{abstract}
\vspace{-3mm}



\received{20 February 2007}
\received[revised]{12 March 2009}
\received[accepted]{5 June 2009}

\maketitle

\vspace{-2mm}
\section{Introduction}

Compiler optimization is critical for unlocking software performance on modern systems~\cite{bacon1994compiler,cummins2024meta}. Production compilers such as LLVM~\cite{lattner2004llvm} and GCC~\cite{stallman2003using} provide a large set of optimization passes~\cite{cummins2022compilergym} that implement program analyses and transformations, such as loop unrolling, instruction scheduling, and register allocation. In practice, developers rely on predefined optimization levels (e.g., \texttt{-O3}, \texttt{-Oz}), which apply fixed pass sequences and parameter settings. However, no single pipeline configuration performs well across programs~\cite{cereda2020collaborative}. As a result, improving compiler performance requires identifying an effective \emph{pass sequence} (known as the phase ordering problem~\cite{ashouri2018survey}) and suitable \emph{parameter settings} for individual passes (e.g., the unrolling factor for the loop unroll pass).

The main barrier for compiler tuning is the scale and structure of the optimization space. Modern compilers include hundreds of passes, yielding a combinatorially large space of possible \textit{pass sequences} and \textit{parameter settings}. Effective configurations are often sparse and highly program-dependent~\cite{cereda2020collaborative}. Search-based autotuning is a common solution~\cite{ansel2014opentuner,ashouri2018survey,cummins2022compilergym} for the problem, as it can explore arbitrary pass combinations without requiring prior training. However, it is computationally expensive. Predictive modeling~\cite{wang2018machine,ashouri2018survey} offers a complementary approach, but it typically requires large training datasets and generalizes poorly across programs, passes, and hardware architectures.

Recent work on large language models (LLMs) offers a new possibility for compiler tuning: generating optimization decisions (e.g., the compiler pass sequence to be used for a given program) directly from program context~\cite{cummins2025llm}. However, existing approaches largely focus on static, deterministic objectives, such as code size~\cite{cummins2024meta,lin2025awarecompiler}, where compiler outcomes are deterministic and directly observable from the code. In contrast, optimizing runtime performance is fundamentally harder. Performance depends on complex microarchitectural interactions, target-specific behavior, and runtime measurements that are often noisy~\cite{dubach2009portable,chen2016robust}. As a result, code-level reasoning alone is insufficient: LLM-generated optimizations may appear reasonable but still yield poor performance on the underlying hardware. The key limitation is the lack of grounded feedback - without compiler-internal signals or runtime evidence, the model cannot reliably evaluate its own decisions.

In this paper, we address this gap with \SystemName, a multi-agent framework that integrates LLMs into the compiler tuning loop through compiler and runtime feedback. \SystemName does not treat the compiler as a black box. Instead, it queries compiler artifacts during compilation, including optimization remarks and LLVM IR snapshots, to expose the effects of transformations. It then iteratively refines optimization decisions using measured runtime performance within a multi-agent framework. This closed-loop design enables the system to diagnose regressions, focus on promising transformations.

We implement \SystemName on top of the LLVM compiler~\cite{lattner2004llvm} and evaluate it on x86-64 and ARM64 platforms. Across a range of benchmarks, \SystemName consistently outperforms strong baselines, including PGO and OpenTuner~\cite{ansel2014opentuner}. Our results show that LLMs can effectively guide compiler optimization when grounded in compiler and runtime evidence, without requiring offline training or task-specific fine-tuning.

This paper makes the following contributions:
\vspace{-1mm}
\begin{itemize}
\item A multi-agent framework that integrates LLMs with compiler-internal signals and runtime feedback for performance optimization.
\item A feedback-driven optimization loop that combines structured pipeline editing, validation, and iterative refinement to improve reliability and efficiency.
\item Empirical results show that inference-only LLMs-based agents can effectively support compiler tuning tasks, including pass ordering and parameter selection.

\end{itemize}

\vspace{-3mm}
\section{Background}
\subsection{Compiler Optimization}

\cparagraph{Compiler pass management.} 
Modern compilers, such as LLVM~\cite{lattner2004llvm}, apply optimizations through a pass manager that schedules a sequence of modular analyses and transformations (e.g., inlining, loop unrolling, and vectorization). Performance depends on both \emph{which} passes run and \emph{how} they are ordered, as well as pass-specific parameters (e.g., unroll factors). These choices interact: applying unrolling before vectorization can expose different IR patterns than doing the reverse, and different parameter settings can change downstream profitability. As architectures diversify and workloads vary, a fixed default pipeline is often not ideal for every program and target, and the resulting search space over pass orderings and configurations grows quickly~\cite{silva2021exploring}.

\cparagraph{Profile-guided optimization.}
PGO is widely used to incorporate runtime behavior into compilation. Common implementations include instrumentation-based PGO~\cite{wicht2014hardware}, which collects explicit edge counts, and sampling-based variants such as AutoFDO~\cite{chen2016autofdo} and CSSPGO~\cite{he2024revamping}. In practice, two issues often limit the gains. First, PGO is sensitive to profile representativeness; when the profiling inputs or environment differ from production, the resulting decisions can overfit and occasionally regress. Second, most deployments keep the overall optimization pipeline largely fixed (e.g., the default \texttt{-O3} structure) and use profiles mainly to steer heuristic decisions and parameters. This leaves less room to explore improvements that require changing the pipeline itself, such as pass reordering or restructuring.

\vspace{-3mm}
\subsection{Compiler AutoTuning}
Iterative autotuning frameworks~\cite{ansel2014opentuner,ren2023javascript} can improve runtime performance, but they often require many compile-and-run evaluations to discover a strong configuration~\cite{zhao2025leveraging}. This cost is driven by the size of the optimization space: modern compilers expose many transformations (e.g., LLVM 17 offers over 100 transformation passes) and analysis passes, and the number of candidate phase orderings grows rapidly even before pass parameters are taken into account. To reduce this search burden, prior work has incorporated machine learning~\cite{wang2018machine,pan2025towards} into compiler optimization, including learned cost models and policy-based selection of optimization sequences. Reinforcement-learning formulations further cast optimization as sequential decision making, and systems such as Autophase~\cite{huang2019autophase}, CompilerGym~\cite{cummins2022compilergym} and Compiler-R1~\cite{pan2025compiler} provide standardized feature interfaces and training environments for learning compiler policies. Despite these advances, such approaches are not directly aligned with practical deployment constraints in our setting. Performance-oriented rewards depend on noisy, hardware- and input-dependent measurements,  and optimization benefits often come from interactions among passes, making them difficult to predict from a fixed feature representation. In addition, learned policies can be hard to interpret and debug, which complicates regression diagnosis under tight evaluation budgets. These factors limit robustness when transferring across workloads, microarchitectures, and compiler versions, where distribution shift is common and retraining or extensive re-tuning is often infeasible.

\begin{table}[t!]
\centering
\caption{Typical two benchmarks used for motivation.}
\label{tab:moti}
\vspace{-3mm}

\footnotesize
\begin{tabular}{l>{\raggedright\arraybackslash}p{0.7\linewidth}}

\toprule
\textbf{Benchmark} & \textbf{Description} \\ 
\midrule
\rowcolor[gray]{.92} \textbf{Qsort} & Implements the well-known divide-and-conquer sorting algorithm. \\ 
\textbf{BitCount} & A collection of algorithms that count the number of set bits in an integer array. \\ 

\bottomrule
\vspace{-8mm}
\end{tabular}
\end{table}

\vspace{-2mm}
\section{Motivation\label{sec:moti}}

As a motivating example, consider pass tuning in LLVM (v17.0.6) to optimize QuickSort (termed as QSort) and BitCount on Intel Core i9 CPU and ARM Cortex-A76. Table ~\ref{tab:moti} lists the benchmarks.

\cparagraph{Setup.}
We conduct the experiments on two hardware platforms: an Intel Core i9 server (x86-64) and a Cortex-A76 embedded device (ARM64). The full platform details are listed in Table~\ref{tab:moti_platforms}. All experiments use Clang/LLVM 17.0.6~\cite{llvm17}. We consider all the individual passes enabled by the LLVM -O3 option.  We compare \SystemName against four baselines: instrumentation-based PGO, AutoFDO, CSSPGO (on x86-64), and a autotuning framework OpenTuner~\cite{ansel2014opentuner}. For OpenTuner, we report the best configuration found within three search iterations (same optimization budget as our approach \SystemName), and its search space is initialized from the default
-O3 pass pipeline. Each configuration is measured five times, and we report geometric-mean speedup relative to \texttt{-O3}. \SystemName uses DeepSeek-V3.2~\cite{deepseek} as the LLM backend.

\begin{algorithm}[t]
\caption{Qsort: a control-flow-heavy motivating example}
\label{alg:qsort}
\small
\begin{algorithmic}[1]
\State \textbf{Parameter:} $\textsc{CUTOFF} \gets 8$

\Function{QSortX}{$base, num, width, comp$}
  \While{$\neg$ \Call{isEmpty}{stack}}
    \If{\Call{size}{$lo, hi, width$} $\le \textsc{CUTOFF}$}
      \Statex \textit{// Hotspot A: small but frequently executed fallback routine}
      \Statex \fbox{\parbox{0.92\linewidth}{
      \textbf{Baseline behavior:} \texttt{-O3}/PGO mainly inline \textsc{ShortSort}, but treat it as low-priority.\\
      \textbf{\SystemName:} applies \emph{inline + loop unroll} because this short routine is repeatedly executed on small partitions.}}
      \State \hlstate{\Call{ShortSort}{$lo, hi, width, comp$}}
      \State \textbf{continue}
    \EndIf

    \While{true}
      \Statex \textit{// Hotspot B: branch-heavy partition scan loop}
      \Statex \fbox{\parbox{0.92\linewidth}{
      \textbf{Baseline behavior:} AutoFDO/CSSPGO and Instr.PGO remains conservative due to branch variance.\\
      \textbf{\SystemName:} still identifies this region as profitable and applies \emph{loop unroll}.}}
      \State \textit{partition scan and pointer movement}
      \If{\hlstate{$higuy < loguy$}}
        \State \textbf{break}
      \EndIf
      \State \Call{Swap}{$loguy, higuy, width$}
    \EndWhile

    \State \textit{push larger partition; continue with the smaller one}
  \EndWhile
\EndFunction
\end{algorithmic}
\end{algorithm}

\begin{algorithm}[t]
\caption{BitCount: a locality-sensitive motivating example}
\label{alg:bitcount_motivation}
\small
\begin{algorithmic}[1]
\Function{BitcountBenchmark}{$dataset\_id, REPEAT\_MAIN$}
    \For{$r \gets 1$ \textbf{to} REPEAT\_MAIN}
        \State $inputs \gets$ \Call{LoadInputs}{$dataset\_id$}
        \State Initialize $S_1, S_2, S_3, S_4, S_5 \gets 0$

        \Statex \textit{// Hotspot: repeated accumulation loop over multiple bit-count kernels}
        \Statex \fbox{\parbox{0.92\linewidth}{
        \textbf{Baseline behavior:} PGO may misclassify the frequently executed \texttt{$S_3$} path as cold, which disrupts code layout and harms L1i locality.\\
        \textbf{\SystemName:} adapts tuning decisions using measured feedback, avoids this misclassification, and preserves better locality.}}

        \For{\textbf{each} $x$ \textbf{in} inputs}
            \State $S_1 \gets S_1 +$ \Call{BitCount\_Shift}{$x$}
            \State $S_2 \gets S_2 +$ \Call{BitCount\_Kernighan}{$x$}
            \State\hlstate{ $S_3 \gets S_3 +$ \Call{BitCount\_Table4}{$x$}}
            \State $S_4 \gets S_4 +$ \Call{BitCount\_Table8}{$x$}
            \State $S_5 \gets S_5 +$ \Call{BitCount\_SWAR}{$x$}
        \EndFor

        \State \Call{PrintChecksums}{$S_1, S_2, S_3, S_4, S_5$}
    \EndFor
\EndFunction
\end{algorithmic}
\end{algorithm}

\begin{figure}[t!]
	\centering
	\subfloat[][X86-64]{\includegraphics[width=0.24\textwidth]{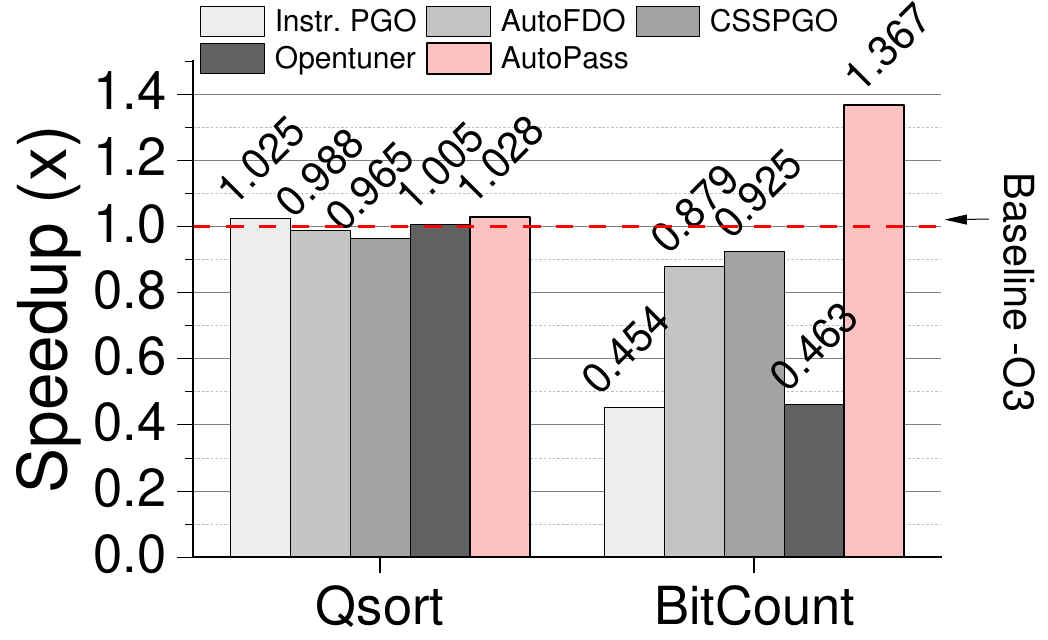}}
	\subfloat[][ARM64 ]{\includegraphics[width=0.244\textwidth]{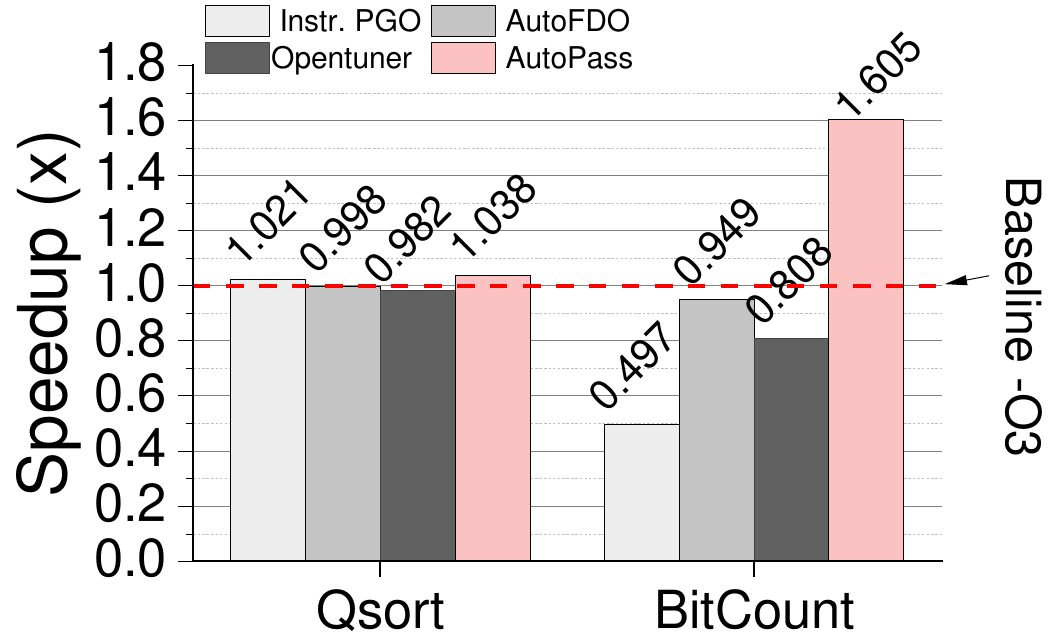}}
	\vspace{-3mm}
	\caption{Performance speedup of Instrumentation-based PGO, AutoFDO, CSSPGO, OpenTuner, and our proposed approach \SystemName, relative to the -O3 baseline. Results are shown for two representative benchmarks on both x86-64 and ARM64 architectures.}
	   \vspace{-8mm}
	\label{fig:moti}
\end{figure}

\cparagraph{Results.}
Algorithms~\ref{alg:qsort} and ~\ref{alg:bitcount_motivation}  list the key divergence points where existing
methods and our approach \SystemName make different optimization decisions on Intel Core i9 server (x86-64)  platform.
In Algorithm~\ref{alg:qsort}, we can see that Qsort’s profitable transformations can be missed even with profile information. The first divergence between \SystemName and the baselines is \textsc{ShortSort} function, which is small but executed frequently. The second divergence point is the partition scan loop (e.g., \textit{if higuy < loguy}), which dominates runtime but exhibits irregular branch behavior. In this case, the PGO baselines tend to remain conservative because high branch-outcome variance weakens local evidence of transformation benefit, whereas \SystemName still identifies the loop as an optimization-critical hotspot.
Algorithm~\ref{alg:bitcount_motivation} highlights a different limitation. BitCount performance is shaped primarily by hot-path identification and instruction-cache locality rather than branch-heavy control flow. In this setting, profile-guided methods can misclassify dominant paths as cold, which degrades layout decisions and leads to their largest regressions. OpenTuner also struggles under the same three-iteration budget, suggesting that small-budget black-box search is often insufficient to reliably discover a strong pipeline. Overall, Figure~\ref{fig:moti} shows that \SystemName achieves the best performance across both benchmarks and platforms, with an average speedup of 1.259$\times$ over \texttt{-O3}.

\cparagraph{Insight.}
The motivating examples highlight two limitations of current optimization workflows. Heuristic-based PGO pipelines remain conservative and miss high-impact regions when evidence about transformation benefit is noisy, while budgeted black-box search often fails to discover the effective pass pipeline under the tight compile-and-run budgets typical of practical deployment. In contrast, \textit{effective compiler optimization requires \emph{context-sensitive} decisions. The compiler must reason about how a code region contributes to the overall algorithm, how transformations interact with control flow and code layout, and how those effects vary across hardware targets.} The emergence of LLMs creates an opportunity to support this kind of reasoning. This motivates our framework, which goes beyond fixed heuristics and revises optimization decisions using richer program evidence and measured performance feedback.

\vspace{-2mm}
\section{Our Approach}

\begin{figure*}[t!]
	\begin{center}
    \includegraphics[width=0.9\textwidth]{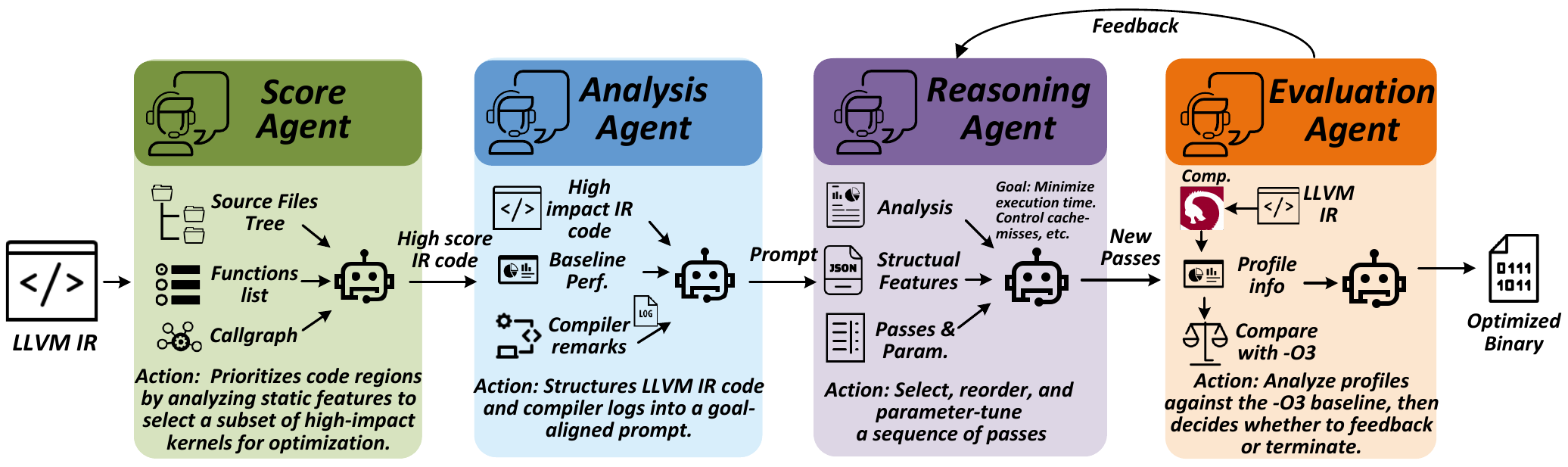}
	\end{center}
 	\vspace{-4mm}
	\caption{Overview of \SystemName, a four-agent LLM-driven LLVM passes generator for performance speedup. }
	 \vspace{-5mm} 
	\label{fig:overview}
\end{figure*}

\SystemName is a multi-agent framework for compiler phase ordering that treats pass-pipeline construction as a guided reasoning process. The goal of \SystemName is to identify optimization pipelines that improve runtime performance under a practical optimization budget. As shown in Figure~\ref{fig:overview}, \SystemName analyzes optimization-relevant program semantics, interprets intermediate compiler artifacts, and iteratively refines pass-pipeline decisions using limited runtime profiling feedback.

\vspace{-2mm}
\subsection{Score Agent: Hotspot Identification}

To address the context-window limitations of LLMs (e.g., 128K tokens for DeepSeek V3.2), the Score Agent first identifies optimization-critical program regions before invoking downstream Analysis agents. Rather than processing the full raw code directly, it scans the source directory to recover the project hierarchy and runs a custom LLVM analysis pass to construct an inter-procedural call graph. It also extracts compact IR-native features, such as basic-block counts and loop counts (Table~\ref{tab:score_features}), without placing the full module into the LLM context. Based on this structural information, the agent assigns each function a priority score and filters out trivial or I/O-bound routines, allowing subsequent agents to focus on high-impact kernels. For selected functions whose IR fits within the context budget, \SystemName then provides the full raw LLVM IR for detailed downstream tasks.

\subsection{Analysis Agent: Feature Extraction and Initial Diagnosis}
The Analysis Agent translates raw LLVM IR into a structured, optimization-relevant state for the Reasoning Agent. It performs two analyses. First, it conducts semantic hint inference by examining symbol names and available metadata to extract high-level cues about the computation, such as whether a function resembles a sorting kernel or a stencil-style loop nest. These cues provide supplementary context that is not directly encoded in standard compiler cost models. Second, it performs remark-guided structural analysis by examining the IR together with compiler diagnostic remarks produced under the baseline \texttt{-O3} pipeline (via \texttt{-Rpass}, \texttt{-Rpass-missed}, and \texttt{-Rpass-analysis}), including signals such as missed vectorization and inlining opportunities. The agent then emits a normalized JSON summary containing (i) semantic hints and (ii) categorized compiler remarks. This structured representation provides an explicit, compiler-grounded basis for downstream policy generation.

\begin{table}[t]
    \centering
            \caption{Static features used for function prioritization} 
            \vspace{-3mm}
    \footnotesize

    \begin{tabular}{ll}
    \toprule
    \textbf{Feature} & \textbf{Description} \\
    \midrule
    \rowcolor[gray]{.92}\texttt{\#Blocks} & Proxy for CFG complexity and instruction-cache pressure. \\
    \texttt{\#Loops} & Targets for high-impact transformations (e.g., unrolling). \\
    \rowcolor[gray]{.92} \texttt{\#Calls} & Reflects call overhead and interprocedural complexity. \\
 \texttt{\#CondBranch} & Captures control-flow irregularity. \\
    \bottomrule
    \end{tabular}
    
   \label{tab:score_features}
            \vspace{-5mm}
\end{table}

\subsection{Reasoning Agent: Core Optimization Decision-Making}
The Reasoning Agent selects and updates the optimization pass pipeline based on compiler evidence and measured runtime behavior. It operates in two stages: an initial proposal step followed by a small number of feedback rounds.
In the first iteration, the agent combines target-specific constraints with the summary produced by the Analysis Agent, and identifies concrete issues such as missed vectorization opportunities or spill-heavy loops. Its goal is to improve runtime performance while avoiding obvious regressions such as increased instruction-cache pressure. Starting from the standard \texttt{-O3} pipeline as a reference, the agent proposes a modified pipeline by selecting, reordering, and parameterizing passes that directly address the identified issues.
In subsequent iterations, the agent incorporates runtime feedback from the previous build-and-run. It compares the new runtime profile with the prior iteration, associates observed latency changes with the corresponding pipeline edits, prunes ineffective transformations, adjusts key parameters (e.g., unroll factors or inlining thresholds), and targets remaining bottlenecks suggested by updated remarks and profiles. After the final round, the system returns the best validated LLVM pass sequence together with a short justification linking major edits to compiler evidence and observed performance changes.

A candidate pipeline produced by the Reasoning Agent is not executed directly. Instead, it is first passed through a deterministic repair-and-validation stage. In our setting, generation errors mainly fall into two categories: malformed pipeline syntax (e.g., missing parentheses, \textit{loop-unroll<unroll-count=4,inline}, where the closing \textbf{>} is missing) and hallucinated pass names (e.g.,\texttt{slp-vector}, which is not a valid LLVM pass name, the right name is \texttt{slp-vectorizer}). We address the former with a script-based syntax checker that detects and completes unmatched parentheses, and the latter by mapping an invalid pass token to the most similar valid pass in the allowed pass set extracted from the initial \texttt{-O3} pipeline. The repaired candidate is then validated by checking: (1) schema correctness of the agent output; (2) membership of all edited passes in the initial \texttt{-O3} pass set; (3) validity of parameter ranges; and (4) successful LLVM compilation and verification. Candidates that still fail any check are rejected before runtime measurement and returned as failed attempts for the next iteration.

\vspace{-3mm}
\subsection{Evaluation Agent: Performance Evaluation and Feedback Loop}

The Evaluation Agent validates each candidate pipeline through compilation, verification, and runtime profiling. At each iteration, it collects execution time, hardware-counter measurements, and updated compiler remarks, and compares the resulting behavior against both the static \texttt{-O3} baseline and the best valid pipeline found so far. Based on these comparisons, it determines whether the candidate improves performance, exposes remaining optimization opportunities, or causes regressions such as increased cache pressure. When a candidate is suboptimal but still informative, the agent summarizes the observed differences and returns them as feedback for the next reasoning round. The loop continues until the iteration budget is exhausted.

Let $t(P)$ denote the mean runtime of pipeline $P$ over three executions, and let $P^\star$ denote the best valid pipeline found so far. For a candidate pipeline $P^{(t)}$, the Evaluation Agent first checks whether it compiles and passes verification. Invalid candidates are rejected immediately, and $P^\star$ is retained. Otherwise, the agent executes the candidate pipeline three times, uses the mean runtime as its measured result, and accepts the candidate only if $t(P^{(t)}) < t(P^\star)$, in which case $P^\star \leftarrow P^{(t)}$. If not, the candidate is rejected and the system rolls back to $P^\star$ for the next iteration. Here, \texttt{-O3} serves as the fixed global reference for reporting speedup, while $P^\star$ serves as the local acceptance reference during iterative search. After the final round, the framework returns $P^\star$ only if it outperforms \texttt{-O3}; otherwise, it falls back to the original \texttt{-O3} pipeline.

\begin{table}[t!]
        		\caption{Evaluation platforms} 
                \label{tab:moti_platforms}
                \vspace{-3mm}
	\begin{center}

		\vspace{-1mm}

		\footnotesize
		 \begin{tabular}{lrrr }
		\toprule
		\textbf{Device} &\textbf{ISA  }&\textbf{CPU }&\textbf{RAM (GB) }\\ 
		\midrule			
		\rowcolor[gray]{.92}  Server& x86-64& Intel Core i9 CPU @ 3.50GHz& 64\\
		 Raspberry Pi 5& ARM64&  Cortex-A76 @ 2.40GHz& 8\\
          \bottomrule

		\end{tabular}

	\end{center}
    		
			 \vspace{-5mm}
\end{table}

\begin{table}[t]
\centering
\caption{Benchmark suites.}\label{tab:bench-suites}
\vspace{-3mm}
\scriptsize
\begin{tabular}{>{\raggedright\arraybackslash}p{0.2\linewidth}p{0.05\linewidth}p{0.62\linewidth}}
\toprule
\textbf{Suite} & \textbf{Cnt.} & \textbf{Role in Evaluation}\\
\hline
\rowcolor[gray]{.92} cBench~\cite{fursin2011milepost} & 31 & General-purpose suite testing whole-program phase-ordering robustness. \\
PolyBench~\cite{Pouchet2012PolyBenchC} & 30 & Loop-intensive kernels targeting vectorization, tiling, and unrolling. \\
\rowcolor[gray]{.92} CoreMark~\cite{EEMBCcoremarkRepo} & 1 & Standard embedded CPU benchmark for fast regression checks. \\
MiniFE~\cite{lin2015assessing} & 1 & HPC sparse linear algebra proxy; stresses memory access patterns. \\
\rowcolor[gray]{.92} LULESH~\cite{LULESHspec} & 1 & Shock hydrodynamics proxy; tests mixed compute-memory interactions. \\
\bottomrule
\end{tabular}
\vspace{-6mm}
\end{table}

\begin{table*}[t!]
\centering
\caption{Performance comparison with rollback policy enabled. \green{Green} cells indicate the best result, and \blue{blue} cells indicate the second-best within each row and platform group. \SystemName achieves the highest average speedup over \texttt{-O3}, outperforming all PGO-based methods and OpenTuner (best in 3 attempts). \texttt{R1} denotes the pipeline produced after the first optimization round, while \texttt{R3} denotes the best performance in three refinement rounds.}
\vspace{-3mm}
\resizebox{\textwidth}{!}{
\begin{tabular}{l|cccccc|ccccc}
\toprule
& \multicolumn{6}{c|}{\textbf{x86-64}} & \multicolumn{5}{c}{\textbf{ARM64}} \\
\textbf{Benchmark} & \textbf{\SystemName (R3)} & \textbf{\SystemName (R1)} & \textbf{Instr.PGO} & \textbf{CSSPGO} & \textbf{AutoFDO} & \textbf{OpenTuner }& \textbf{\SystemName (R3)} & \textbf{\SystemName (R1)} & \textbf{Instr.PGO} & \textbf{AutoFDO} & \textbf{OpenTuner }\\
\midrule

cBench
& \best{1.059}
& \second{1.046}
& 1.037
& 1.018
& 1.012
& 1.035
& \best{1.111}
& 1.055
& 1.037
& 1.028
& \second{1.088}
\\

PolyBench
& \best{1.009}& 1.005
& 1.001
& 1.006
& 1.006
& \best{1.009}& \best{1.149}
& \second{1.129}
& 1.011
& 1.012
& 1.012
\\

CoreMark
& \best{1.137}
& \second{1.117}
& 1.004
& 1.063
& 1.005
& 1.093
& \best{1.091}
& 1.006
& \best{1.091}
& 1.083
& 1.047
\\

MiniFE
& \best{1.008}& 1.000
& 1.003
& \second{1.006}
& 1.001
& 1.000
& \best{1.068}
& \second{1.039}
& 1.023
& 1.000
& 1.004
\\

LULESH
& \best{1.102}
& 1.089
& \second{1.101}
& 1.077
& 1.004
& 1.066
& \best{1.046}& 1.004& \second{1.040}& 1.020
& 1.010
\\

\bottomrule
\end{tabular}
}
\label{tab:overall}
\vspace{-3mm}
\end{table*}

\vspace{-3mm}
\section{Experimental \label{sec:setup}}

\subsection{Research Questions}
To evaluate the effectiveness of \SystemName, we conduct experiments to answer the following research questions (\textbf{RQs}):
\begin{itemize}
    \item \textbf{RQ1:} Under a strictly constrained budget of target-side executions, can \SystemName deliver greater and more stable execution speedups than established traditional and search-based compiler tuning baselines?
    \item \textbf{RQ2:} How much does iterative feedback contribute beyond one-shot optimization?
    \item \textbf{RQ3:} Does \SystemName adapt its optimization behavior across architectures instead of using a one-size-fits-all policy?
    \item \textbf{RQ4:} Can the Score Agent identify optimization-critical functions more effectively than standard PGO-based hot-function selection?
    \item \textbf{RQ5:} Which components of the multi-agent design are most critical for effectiveness and robustness?
    \item \textbf{RQ6:} How does grounding help make \SystemName's optimization decisions interpretable and diagnosable?
\end{itemize}

\vspace{-3mm}
\subsection{Experimental Setup}
\vspace{-1mm}
\cparagraph{Hardware and Software.}
We evaluate \SystemName on two hardware architectures: a server-grade x86-64 workstation (Intel Core i9-11900K) and an embedded ARM64 edge device (Raspberry Pi 5, Cortex-A76), as detailed in Table~\ref{tab:moti_platforms}. The systems run Ubuntu 20.04 LTS To ensure stable timing, we disable dynamic frequency scaling (Turbo Boost) on the server platform. \SystemName is built as a multi-agent workflow using CrewAI~\cite{crewai_docs2025}, with DeepSeek-V3.2 as the main reasoning backend (other LLM backends are evaluated in Section~\ref{sec:general}).

\cparagraph{Compiler and passes.}
All experiments are conducted using LLVM/Clang 17.0.6~\cite{llvm17} with the New Pass Manager. Our evaluation considers 74 LLVM optimization passes and allows compiler sequences of up to 107 passes. 

\cparagraph{Baselines.}
We compare \SystemName against four baselines: Instrumented PGO, CSSPGO (x86 only), AutoFDO, and the representative search-based autotuner OpenTuner. To ensure a fair comparison, OpenTuner is assigned the same optimization budget as \SystemName, i.e., three iterations, and its search is initialized from the default \texttt{-O3} pass pipeline.
To meet the practical deployment requirement, we employ a \texttt{Rollback Mechanism} for all methods. If we detect a speedup ratio $< 1.0$ (performance degradation), the system automatically discards the candidate and reverts to the \texttt{-O3} baseline.

We do not include certain learned-policy approaches in the comparison because either their trained model weights (such as ACPO~\cite{ashouri2023acpo}) are not publicly available for reproduction, or their primary optimization objective differs from ours (e.g., Autophase~\cite{huang2019autophase} and CompilerGym~\cite{cummins2022compilergym} primarily target code size reduction rather than execution speed).

\cparagraph{Metrics.}
We report performance as speedup over the \texttt{-O3} baseline. For each benchmark, the \texttt{-O3} binary and the optimized binary are each exeted 5 times, and their mean runtimes are used to compute $Speedup = \frac{T_{O3}}{T_{\text{opt}}}$,
where $T_{O3}$ and $T_{\text{opt}}$ denote the mean runtime of the \texttt{-O3} and optimized binaries, respectively. For benchmark suites, we aggregate benchmark-level speedups using the geometric mean. 

\cparagraph{Benchmarks.}
To evaluate the generalization capability of \SystemName, we employ a diverse suite of 5 standard benchmarks in compiler optimization spanning embedded systems, scientific computing, and synthetic stress tests, comprising a total of 64 distinct workloads (Table~\ref{tab:bench-suites} lists the details).

\vspace{-3mm}
\section{Evaluation\label{sec:result}}
\vspace{-1mm}
\subsection{Overall Results (RQ1)}

Table~\ref{tab:overall} reports the speedup of \SystemName and several representative baselines over \texttt{-O3} on five benchmark suites across x86-64 and ARM64. Overall, \SystemName (R3) achieves the strongest performance in 9 out of 10 platform--suite settings, indicating that the proposed grounded multi-agent workflow is effective across both server-class and embedded targets. 
On x86-64, \SystemName (R3) delivers strong improvements on CoreMark ($1.137\times$) and LULESH ($1.102\times$). On ARM64, \SystemName (R3) delivers an average speedup of $1.117\times$ over -O3.
\vspace{-1mm}
\begin{tcolorbox}[colback=olive!20,colframe=black,left=1mm,right=1mm,top=1mm,bottom=1mm,arc=2pt,boxrule=0.6pt]
\textbf{\textit{RQ1:}} \textit{Under a strict on-device iteration budget, AUTOPASS achieves the highest overall execution speedup across all evaluated suites and platforms, outperforming industrial FDO variants and budget-constrained search baselines. Through constrained pipeline editing and evidence-guided refinement, it delivers the most reliable performance gains while maintaining strict baseline stability.}
\end{tcolorbox}
\vspace{-1mm}

Compared to PGO-based methods, \SystemName is more consistent across workloads. Instr.PGO, CSSPGO, and AutoFDO provide moderate improvements in selected cases, but their gains are often close to parity with \texttt{-O3} and vary substantially across suites. For example, Instr.PGO performs well on LULESH, but is much less effective on CoreMark and PolyBench. This suggests that profile-guided methods remain conservative in their optimization choices and heavily depend on the quality of the collected profiling data. In contrast, \SystemName adapts the pass pipeline using compiler diagnostics and measured runtime behavior, which allows it to make optimization decisions from richer evidence than profile-guided methods alone. OpenTuner, as a representative autotuning method, is less stable under a limited search budget. By contrast, \SystemName starts from the compiler-supported \texttt{-O3} pipeline, performs constrained edits, and refines them using execution feedback, enabling it to discover stronger pipelines with fewer attempts. From a deployment perspective, \SystemName is also more practical. Unlike instrumented PGO, which inserts profiling instructions and perturbs runtime behavior during data collection, and unlike AutoFDO, which still requires substantial profiling effort, \SystemName can keep all agent reasoning in the cloud and requires only compiler artifacts and runtime measurements from the target platform. This reduces on-device overhead and makes the framework more suitable for deployment-constrained environments.

\begin{table}[t!]
\centering
\caption{Performance comparison without rollback policy on the cBench dataset. Speedups are reported relative to \texttt{-O3}.}
\vspace{-3mm}
\footnotesize
\begin{tabular}{>{\raggedright\arraybackslash}p{0.08\linewidth}lr>{\raggedleft\arraybackslash}p{0.06\linewidth}>{\raggedleft\arraybackslash}p{0.06\linewidth}rr}
\toprule
\textbf{Platform} & \textbf{Method} & \textbf{Geo. Mean} & Wins ($\geq$1.0) & Losses ($<$1.0) & Max. & Min.\\
\midrule

\multirow{6}{*}{\textbf{x86-64}}
& \SystemName (R3) & \best{1.040} & 25 & 6 & 1.366 & 0.784 \\
& \SystemName (R1) & \second{1.010} & 18 & 13 & 1.275 & 0.753 \\
& Instr. PGO & 0.997 & 21 & 10 & 1.186 & \textbf{0.454} \\
& CSSPGO & 0.993 & 16 & 15 & 1.151 & 0.801 \\
& AutoFDO & 0.987 & 14 & 17 & 1.095 & 0.748 \\
& OpenTuner (3 iter.) & 0.991 & 19 & 12 & 1.165 & 0.544 \\
& OpenTuner (500 iter.)  & \oracle{1.057} & 23 & 8 & 1.644 & 0.861 \\
\midrule

\multirow{5}{*}{\textbf{ARM64}}
& \SystemName (R3) & \best{1.109} & 27 & 4 & 2.040 & 0.961 \\
& \SystemName (R1) & 1.004 & 15 & 16 & 2.028 & 0.728 \\
& Instr. PGO & 0.999 & 21 & 10 & 1.156 & 0.497 \\
& AutoFDO & 1.019 & 21 & 10 & 1.110 & 0.882 \\
& OpenTuner (3 iter.) & \second{1.079} & 26 & 5 & 2.622 & 0.769\\
& OpenTuner (500 iter.) & \oracle{1.126} & 26 & 5 & 2.756 & 0.844\\
\bottomrule
\end{tabular}

\label{tab:performance_wo_rb}
\vspace{-5mm}
\end{table}

\begin{figure}
	\begin{center}
    \includegraphics[width=0.26\textwidth]{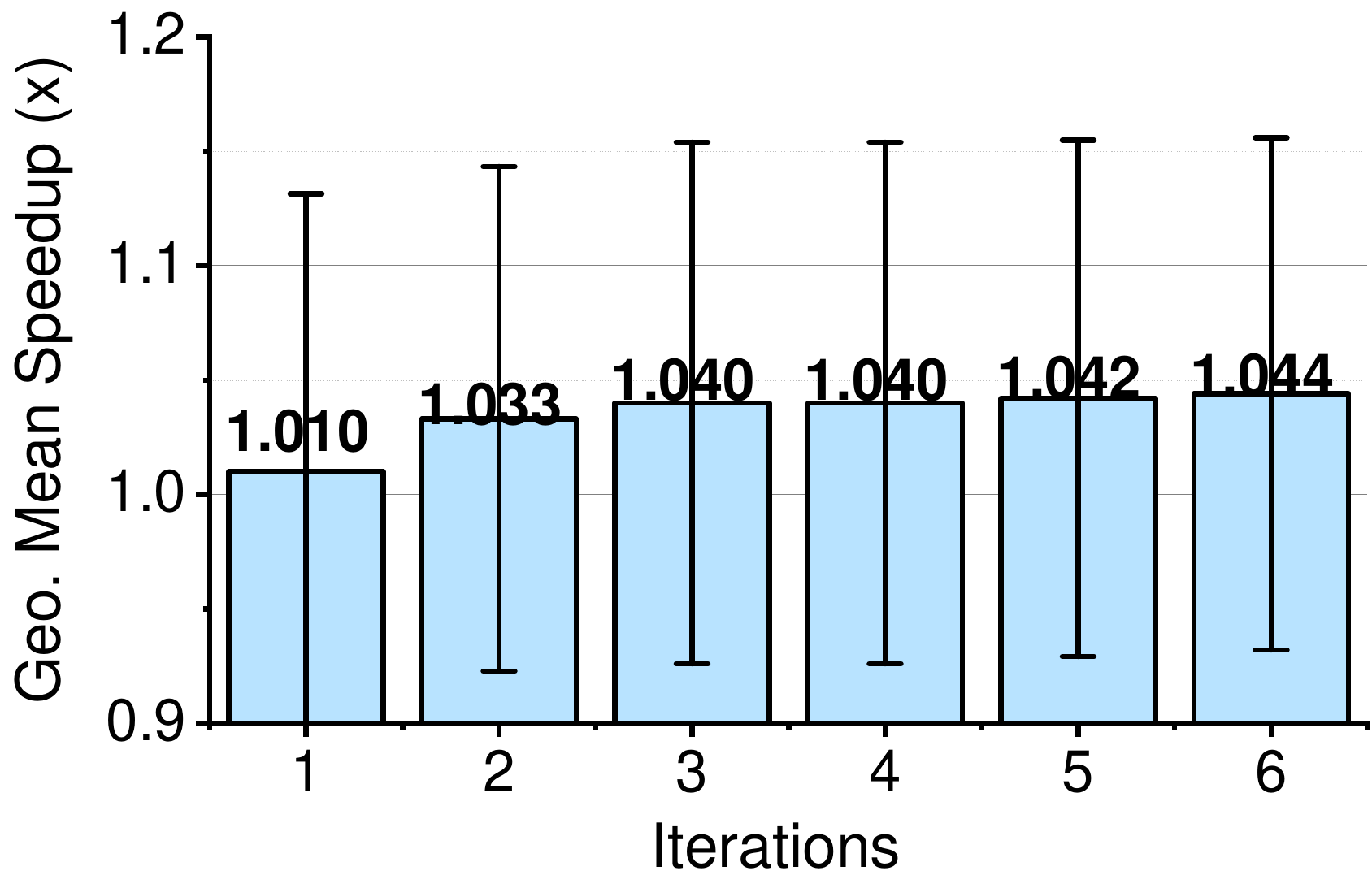}
	\end{center}
 	\vspace{-4mm}
	\caption{Geomean speedup over \texttt{-O3} across six optimization iterations on cBench (x86-64) without rollback policy.}
	 \vspace{-7mm} 
	\label{fig:iter}
\end{figure}

\vspace{-2mm}
\subsection{Performance Without Roll-Back Policy (RQ2) }
Table~\ref{tab:performance_wo_rb} reports the performance of different approaches \textit{without} the ``Safe Rollback" policy on the cBench dataset.
\subsubsection{Performance without roll-back policy}
The results show that \SystemName consistently outperforms all baselines across both platforms. On the server platform (x86-64), it achieves a geometric mean speedup of 1.040x (Max 1.366x) with only 6 regressions (details of the failure cases are available at \url{https://anonymous.4open.science/r/AutoPass-2C75}), outperforming all three PGO baselines (Instrumented, CSSPGO, AutoFDO), which yield a geometric mean speedup below 1.0$\times$, with severe degradation in worst-case scenarios (Min $0.454\times$).  This confirms that rigid heuristic-based profiling may misalign with runtime behavior, causing regressions that blind application of PGO cannot prevent. On ARM64, \SystemName delivers a geometric mean of 1.109$\times$ and a peak speedup of 2.040$\times$. The greater improvement on ARM64 highlights that \SystemName is able to exploit the conservative nature of LLVM's default pipeline on the embedded platform (ARM64). While standard -O3 heuristics often avoid aggressive unrolling or vectorization to strictly manage code size, \SystemName leverages its hybrid reasoning to safely deploy these optimizations and bridges the gap between conservative defaults and hardware capability. For OpenTuner (best in 3 iterations), it proves effective on ARM64 (Mean $1.079\times$), its performance is characterized by extreme volatility. It achieves the highest single-benchmark ($2.622\times$) but also suffers from deep regressions (Min $0.769\times$), typical of blind evolutionary search. We additionally report OpenTuner with 500 iterations as a high-budget search reference. Although this setting achieves the highest average speedup, it still incurs more failure cases than \SystemName, suggesting that a higher search budget improves peak optimization quality but does not guarantee the same level of robustness.

\begin{figure*}[t!]
	\centering
	\subfloat[][Loop Unroll]{\includegraphics[width=0.2\textwidth]{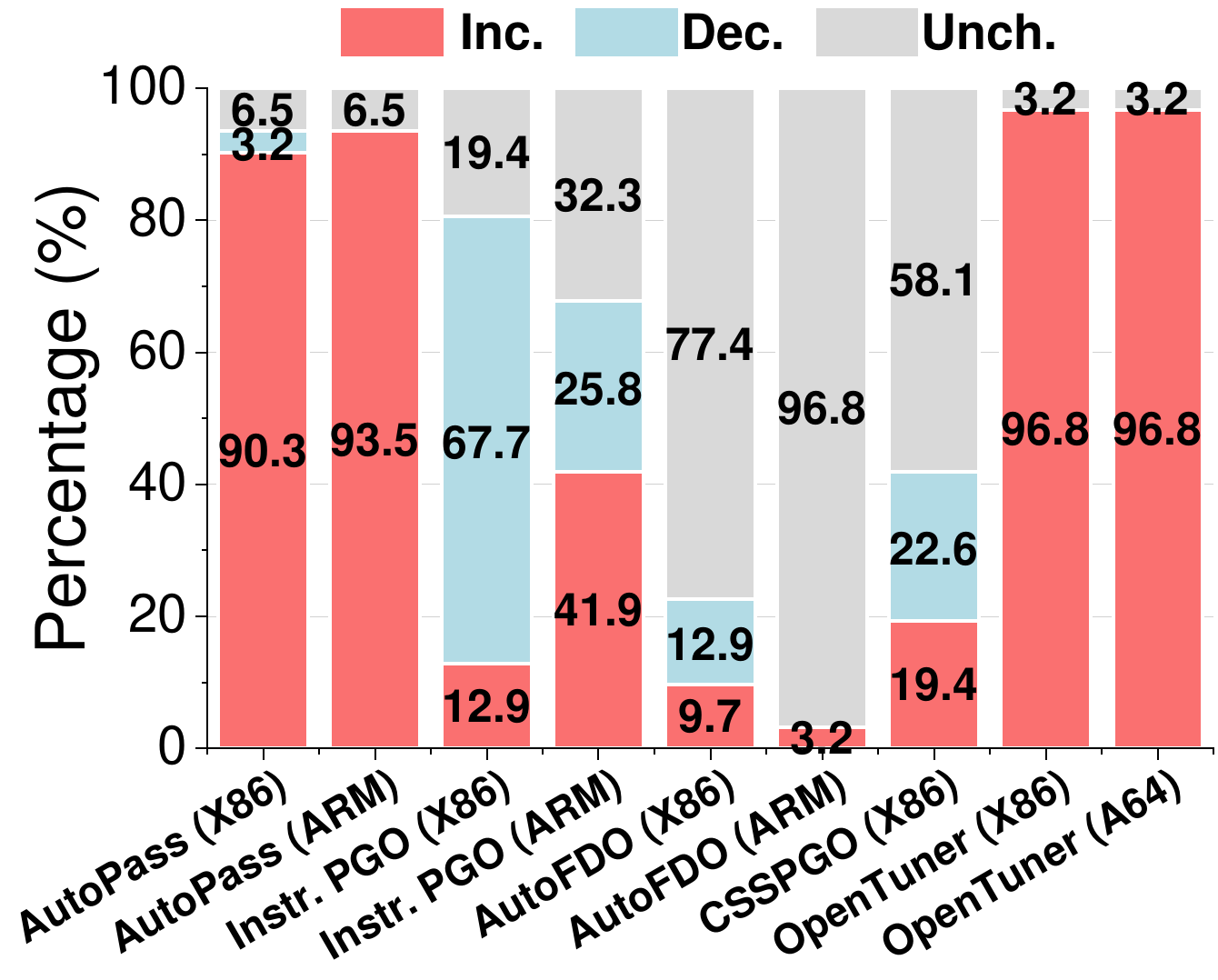}}
    \subfloat[][SLP Vectorizer]{\includegraphics[width=0.2\textwidth]{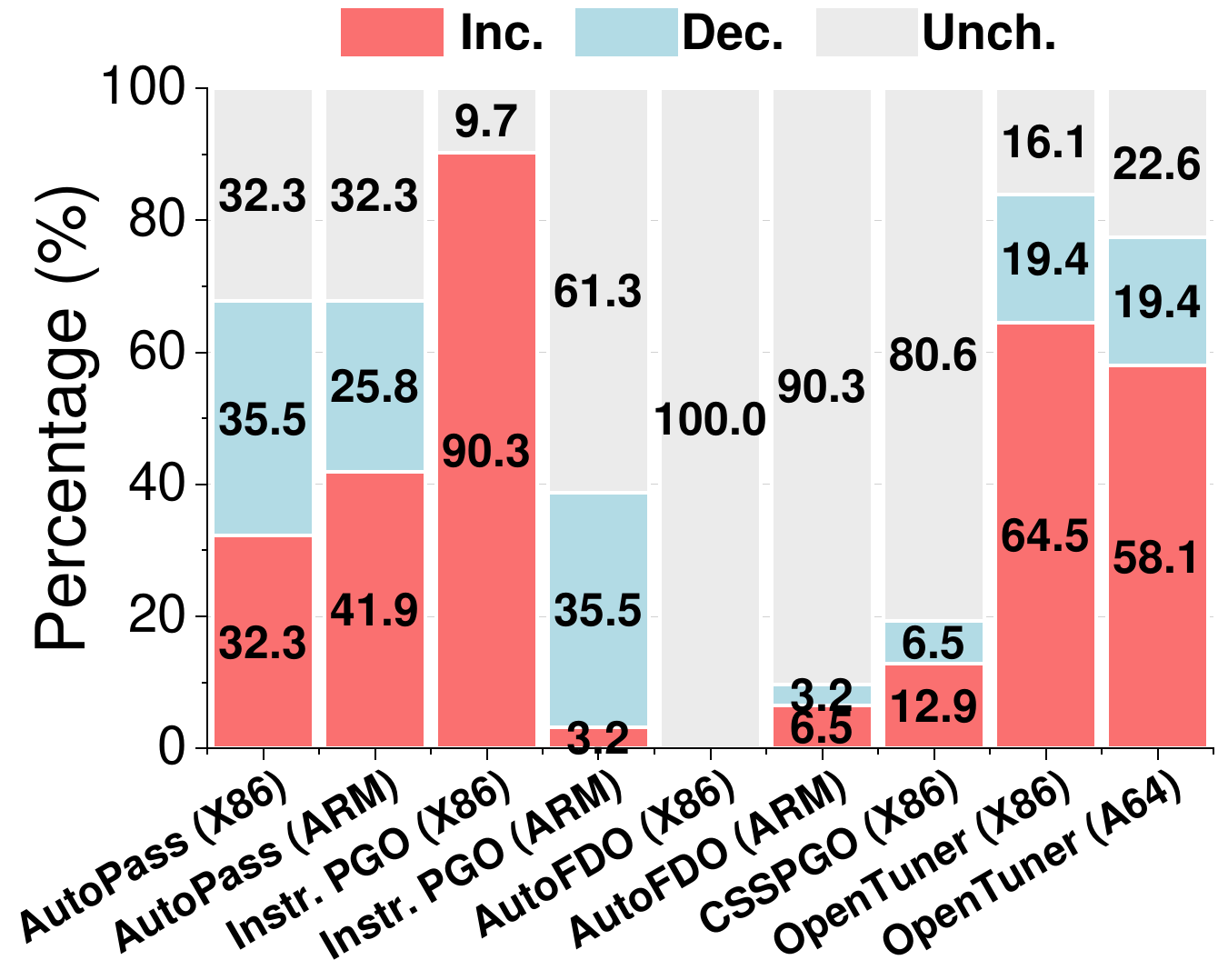}}
         \subfloat[][LICM]{\includegraphics[width=0.2\textwidth]{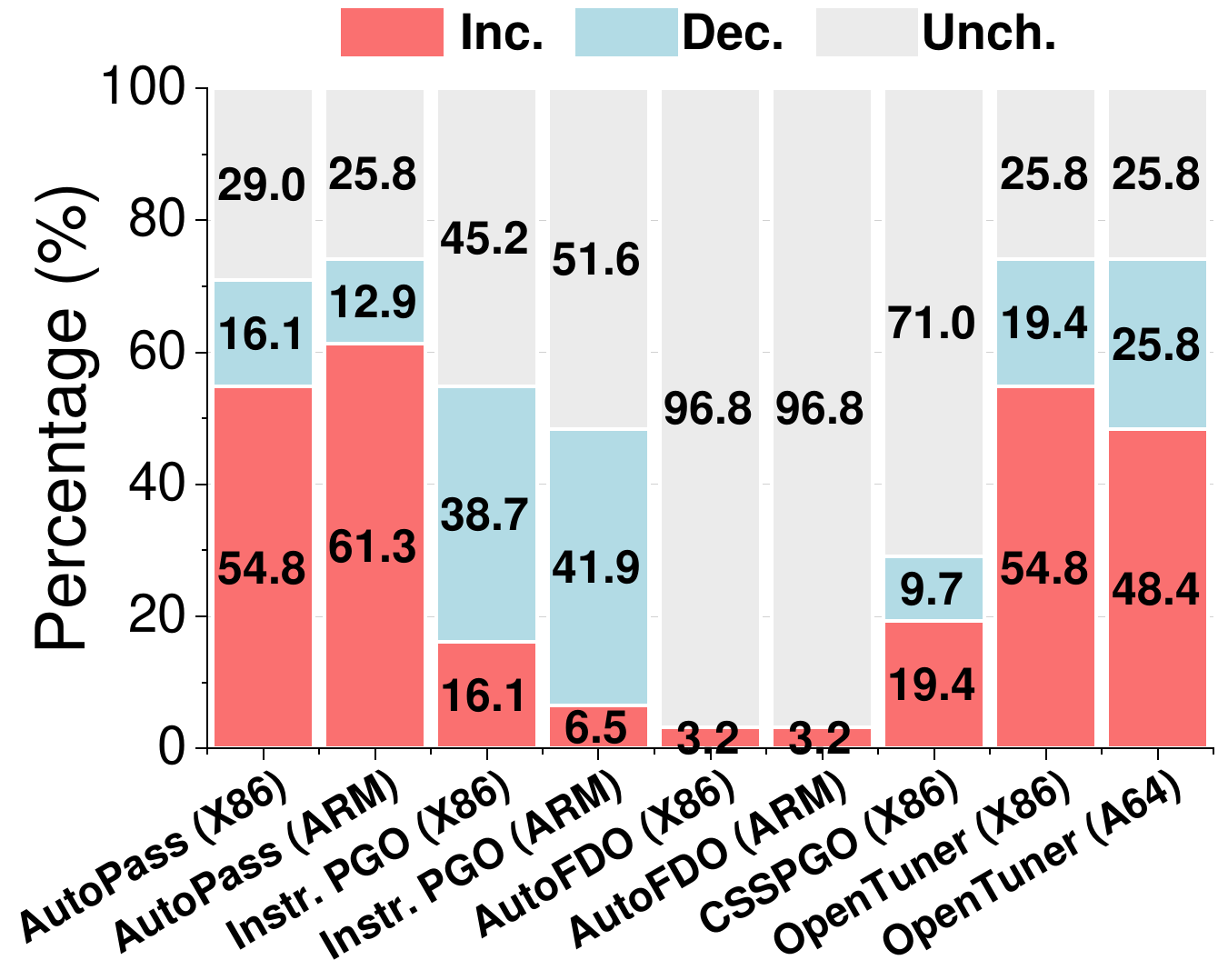}}
    \subfloat[][Inlining]{\includegraphics[width=0.2\textwidth]{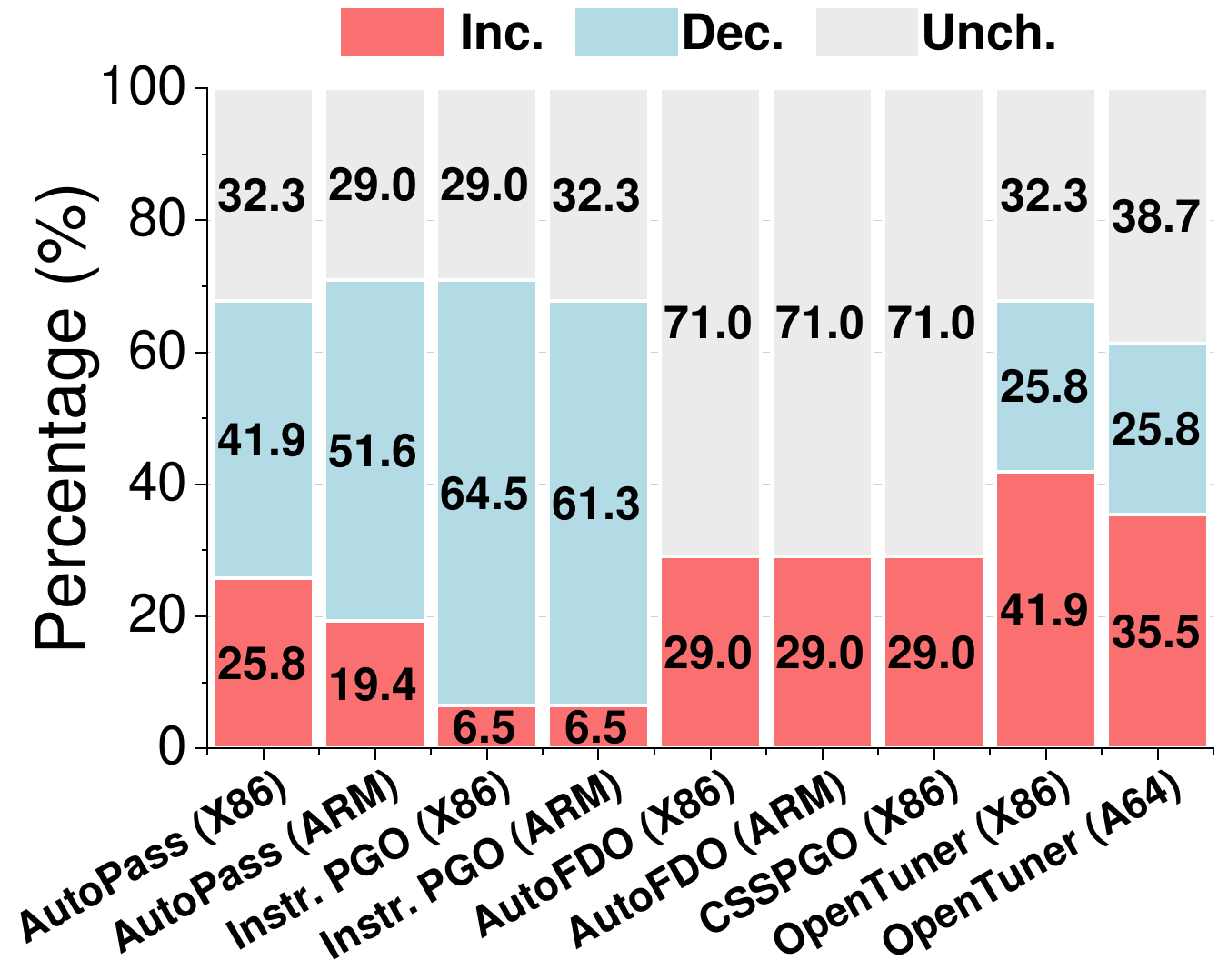}}
    \subfloat[][Tail Call Elim.]{\includegraphics[width=0.2\textwidth]{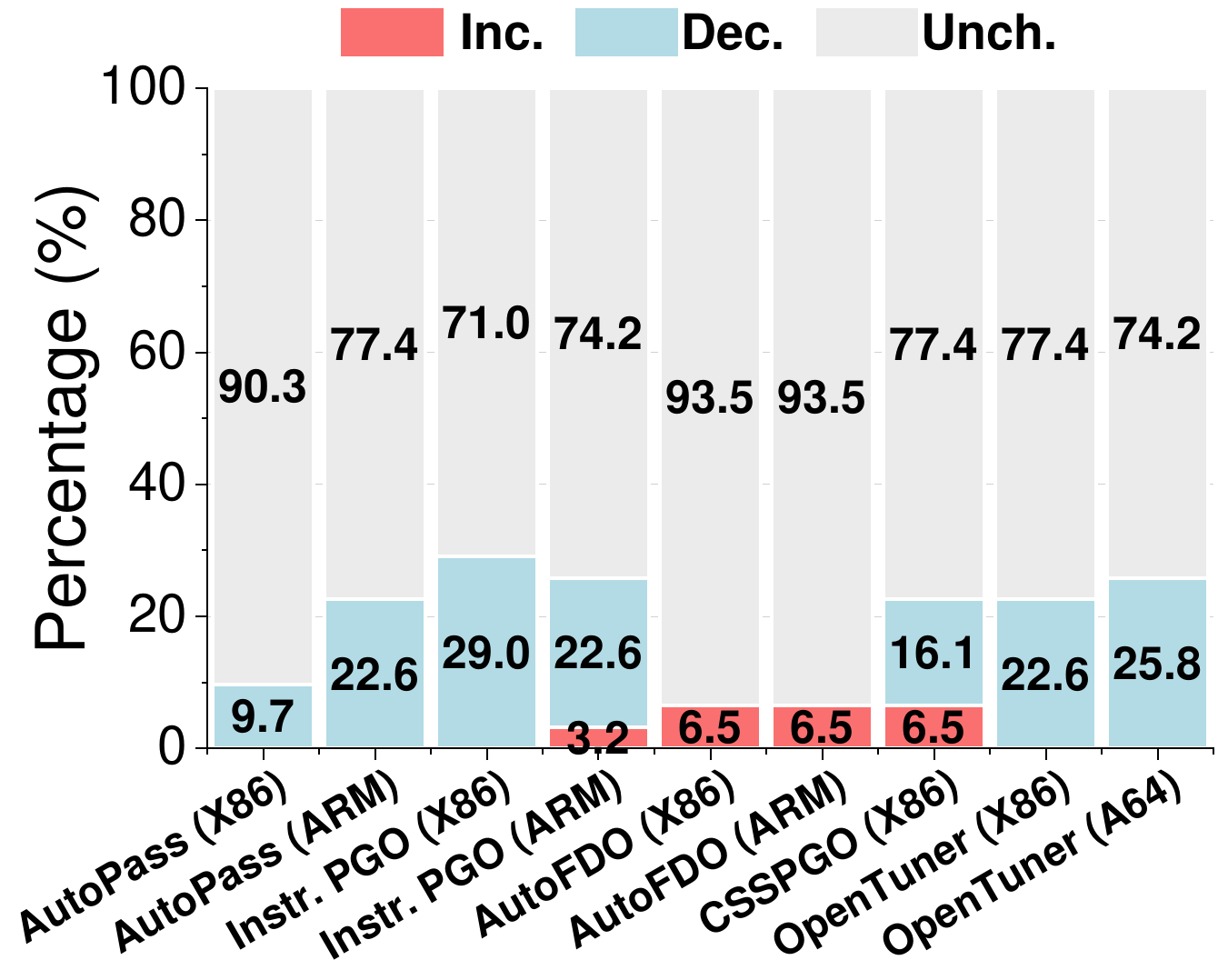}}
       \vspace{-4mm}
	\caption{Percentage of cBench benchmarks for which \SystemName changes the number of effective applications of each pass relative to \texttt{-O3}, classified as increased (Inc.), decreased (Dec.), or unchanged (Unch.), on x86-64 and ARM64.}
	   \vspace{-4mm}
	\label{fig:passfreq}
\end{figure*}
\vspace{-2mm}
\begin{tcolorbox}[colback=olive!20,colframe=black,arc=3pt,boxrule=0.8pt]
\textbf{\textit{RQ2:}} \textit{Iterative feedback contributes substantially beyond one-shot optimization. Compared with R1, AutoPass (R3) achieves higher geometric-mean speedup, reduces regressions from 13 to 6 on x86-64, confirming that feedback-driven refinement is critical for both effectiveness and stability.}
\end{tcolorbox}
\vspace{-2mm}

\vspace{-2mm}
\subsubsection{One-Shot vs. Iterative Efficacy}
The performance gap between single-shot (R1) and iterative (best in three iterations termed as R3) inference underscores the critical role of feedback in stabilizing LLM-driven optimization. While R1 identifies beneficial transformations, it operates as a `cold start' optimization without historical context, evidenced by 13 regressions on the server platform (x86-64) and a marginal geometric mean of 1.010$\times$.
However, by best performance in R3, the \SystemName successfully prunes these ineffective strategies, reducing the loss count to just 6 while increasing the win count from 18 to 25. This improvement proves that the Evaluation Agent functions not just as a filter, but as a constructive critic that guides the system toward valid optimization subspaces. 

\subsubsection{Search Convergence Analysis}
Figure~\ref{fig:iter} presents the geometric mean speedup across six optimization iterations on cBench. While one-shot reasoning yields only marginal gains (1.010$\times$), the second iteration drives a sharp increase to 1.033$\times$, confirming the efficacy of feedback-driven correction. Performance effectively saturates at Iteration 3 (1.040$\times$), with subsequent rounds yielding negligible improvement (max 1.044$\times$). Consequently, we set the termination threshold at three iterations to balance optimization quality with computational efficiency.
\vspace{-1mm}

\vspace{-3mm}
\subsection{Architecture-Aware Optimization Behavior (RQ3)}
\subsubsection{Analysis of Optimization Coverage}
To diagnose performance divergence, we quantify the optimization coverage of five key compiler passes relative to the \texttt{-O3} baseline. For each benchmark, we count how many times a given pass is reported as effective in the compiler optimization remarks, and use this count as a proxy for how many optimization opportunities that pass actually affects. A benchmark is then classified as \emph{Increased}, \emph{Decreased}, or \emph{Unchanged} depending on whether the pass is effective more often, less often, or equally often as under \texttt{-O3}. Figure~\ref{fig:passfreq} summarizes the resulting distribution across the cBench suite.

\cparagraph{Consistent Expansion of Instruction-Level Parallelism (ILP) in \SystemName.}
Across both architectures, \SystemName adopts a unified strategy of expanding instruction-level parallelism. It increases loop-unrolling coverage in 90.3\% of x86-64 and 93.5\% of ARM64 benchmarks, indicating that the agent frequently identifies more loop regions as worth unrolling than the default LLVM cost model does. This increase is accompanied by higher LICM  coverage in 55--61\% of programs. Together, these patterns suggest a compensatory strategy: when more loops are unrolled, \SystemName also increases the amount of loop-invariant code hoisted out of those loops, reducing repeated work and mitigating the additional pressure introduced by larger loop bodies.

\cparagraph{Architecture-Aware Vectorization.}
\SystemName also shows architectural sensitivity in its vectorization behavior. On x86-64, it increases SLP vectorization coverage in 32.3\% of cases, decreases it in 35.5\%, and leaves it unchanged in the remaining benchmarks. On ARM64, it increases SLP coverage more often, in 41.9\% of cases, while decreasing it in 25.8\% and leaving the rest unchanged. This pattern suggests that \SystemName is more willing to apply aggressive vectorization on ARM64, while adopting a more balanced strategy on x86-64. Although both platforms support SIMD, vectorization profitability remains target-dependent, and \SystemName adjusts its behavior accordingly.

\begin{table}[t!]
    \centering
        \caption{Edit similarity analysis of optimization pass sequences (excluding parameter settings). We report the geometric mean edit similarity between \SystemName-generated pipelines and the default \texttt{-O3} baseline for x86-64 and ARM64, as well as the cross-architectural similarity between the generated pipelines for both platforms in the cBench dataset.}
            \label{tab:es}
            \vspace{-3mm}
    \footnotesize
    \begin{tabular}{l|>{\centering\arraybackslash}p{0.16\linewidth}>{\centering\arraybackslash}p{0.18\linewidth}|>{\centering\arraybackslash}p{0.16\linewidth}}
\toprule
         \textbf{Metric}& \SystemName vs -O3 (x86)& \SystemName vs -O3 (ARM)& \SystemName x86 vs ARM\\
        \midrule
         Geo. Mean with SD.& 0.943$\pm$0.050& 0.930$\pm$0.042& 0.917$\pm$0.046\\
         \midrule
         Min Value& 0.768 & 0.821 & 0.800\\
         \midrule
         Max Value& 1.000& 1.000& 0.988\\
         \bottomrule
    \end{tabular}
    \vspace{-6mm}

\end{table}

\cparagraph{Limitations of Traditional PGO.}
Instrumented PGO exhibits a bias toward local, block-level optimization at the expense of broader loop restructuring on x86-64. In particular, it tends to increase SLP vectorization coverage while reducing loop-unrolling coverage. This trade-off can hurt loop-intensive workloads, since local vectorization alone does not necessarily preserve the regular execution structure needed for efficient iteration. In contrast, sampling-based methods such as AutoFDO and CSSPGO show much stronger structural rigidity, as they largely preserve the optimization coverage pattern of the fixed \texttt{-O3} pipeline. For example, for Tail Call Elimination, over 93\% of benchmarks show unchanged coverage across both architectures. This suggests that these methods mainly refine heuristic decisions within the existing \texttt{-O3} structure, rather than reshaping which program regions are transformed.

\cparagraph{Stochastic Aggression of Evolutionary Search.}
OpenTuner applies aggressive transformations in a much less selective manner. It increases loop-unrolling coverage in 96.8\% of benchmarks on both architectures, indicating a broad tendency to expand loops regardless of workload structure. It also shows weaker architectural sensitivity than \SystemName. As shown in Figure~\ref{fig:passfreq}(b), OpenTuner increases SLP vectorization coverage on x86-64 in 64.5\% of cases, whereas \SystemName does so in only 32.3\%. This suggests that OpenTuner tends to push vectorization more uniformly, even when the target architecture makes such transformations less attractive.
\vspace{-1mm}
\begin{tcolorbox}[colback=olive!20,colframe=black,left=1mm,right=1mm,top=1mm,bottom=1mm,arc=2pt,boxrule=0.6pt]
\textbf{\textit{RQ3:}} \textit{\SystemName is architecture-aware, adapting pass behavior and pipeline structure across hardware targets rather than applying a uniform policy.}
\end{tcolorbox}
\vspace{-1mm}

\begin{figure}
	\begin{center}
    \includegraphics[width=0.45\textwidth]{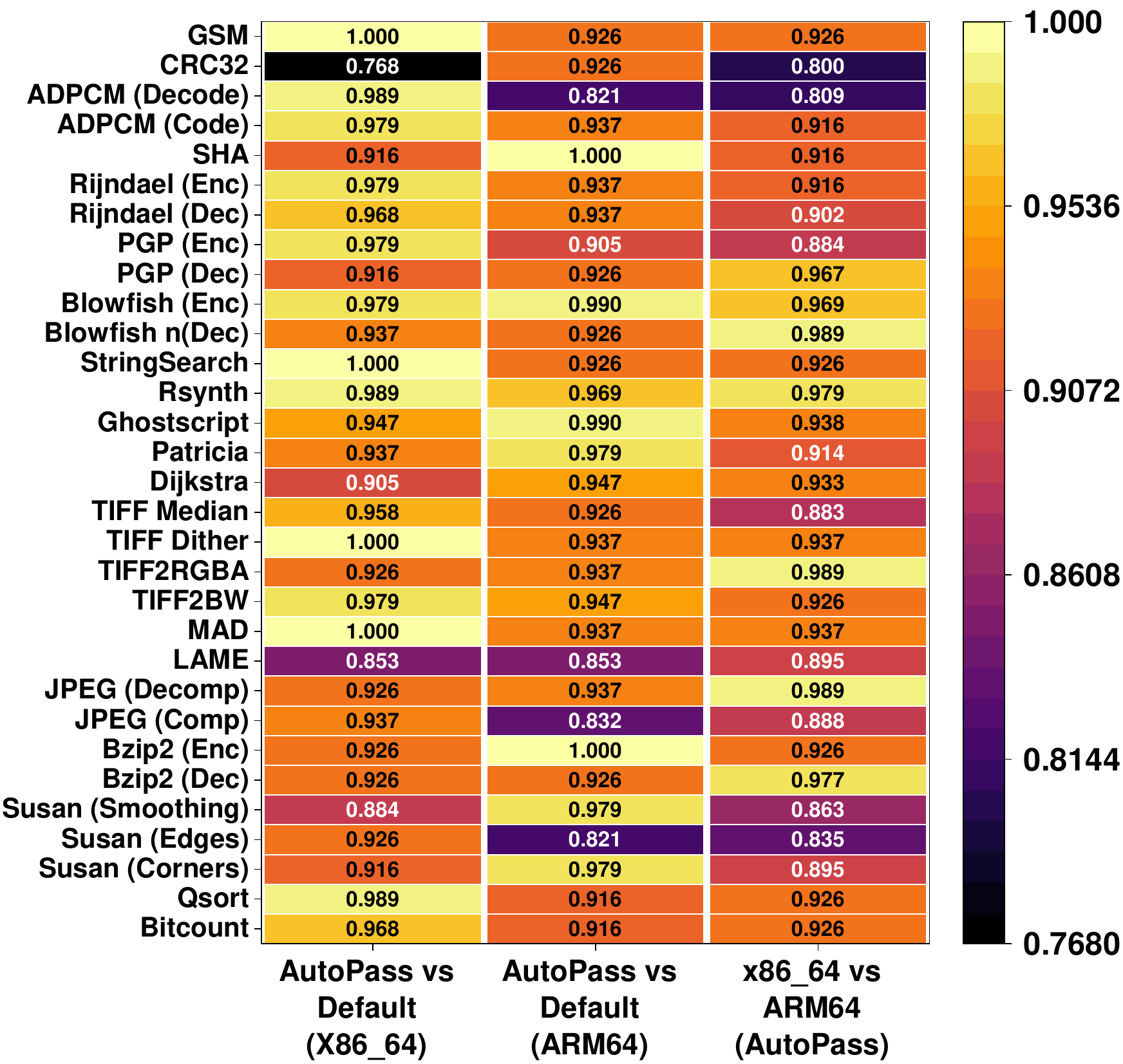}
	\end{center}
 	\vspace{-3mm}
	\caption{Heatmap of $ES$ for individual cBench benchmarks. }
	 \vspace{-3mm}
	\label{fig:heatmap}
\end{figure}

	 \vspace{-2mm}
\subsubsection{Quantitative Analysis of Pipeline Topology}
To quantify structural divergence, we employ Edit Similarity ($ES$)~\cite{armengolestapé2024sladeportablesmalllanguage} to measure the topological distance between the pass sequence generated by \SystemName and the default -O3 baseline (Table~\ref{tab:es} lists the results). Figure~\ref{fig:heatmap} visualizes these deviations across 31 benchmarks on x86-64 and ARM64. The results indicate that \SystemName works as a selective, architecture-aware optimizer.

\cparagraph{Adaptive Pipeline Strategy.} We can observe that \SystemName employs a selective optimization strategy rather than random exploration. For benchmarks like \texttt{StringSearch} (x86) and \texttt{SHA} (ARM), high similarity scores ($ES = 1.0$) indicate that the system retains the default pipeline structure when the baseline is already effective. In contrast, for workloads with distinct bottlenecks like \texttt{CRC32} (x86, $ES=0.768$) and \texttt{JPEG} (ARM, $ES=0.832$), the system significantly alters the pass order.

\cparagraph{Cross-Platform Orthogonality.} The $ES$ between \SystemName's x86 and ARM pipelines averages only 0.917, indicating target-specific pipelines. For the computationally dense CRC32 benchmark, the agent aggressively reorders the x86 pipeline ($ES=0.76$) to saturate wide issue slots, while retaining a conservative strategy on ARM ($ES=0.92$) to prevent detrimental code expansion. This distinction highlights the system's ability to navigate hardware constraints such as register pressure and instruction density.

\begin{table}[t]
\centering
\caption{Average speedup and function-selection overlap under different Top-k settings. Overlap is defined as the average intersection ratio between the function sets selected by PGO and by the Score Agent.}
\label{tab:avg_speedup_overlap}
\vspace{-3mm}
\footnotesize
\begin{tabular}{lcccc}
\toprule
\textbf{Function}& \textbf{OpenTuner (500 iter.)} & \textbf{PGO-hot} & \textbf{Score Agent} & \textbf{ Overlap}\\
\midrule
Top-5  & 1.0335 & 1.0277 & 1.0270 & 32.2\%\\
Top-10 & 1.0335 & 1.0277 & 1.0333 & 35.5\%\\
Top-20 & 1.0335 & 1.0219 & 1.0221 & 31.6\%\\
\bottomrule
\end{tabular}
\vspace{-6mm}
\end{table}

\subsection{The Impact of Score Agent (RQ4)}
To study the impact of the Score Agent, we compare two function-selection strategies under different Top-\(k\) settings: instrumented PGO-based hot-function selection and Score-Agent-guided function selection. As a full-program reference, we also report OpenTuner (500 iter.) as an approximate upper bound, since it searches over the entire program rather than selecting hot functions. Table~\ref{tab:avg_speedup_overlap} summarizes the average speedup and the average overlap between the functions selected by PGO and by the Score Agent. We can see that the Score Agent consistently improves over or matches PGO-based selection, and its best result appears in the Top-10 setting. This result suggests that accurate ranking of a modest number of optimization-critical functions is sufficient to recover nearly all the achievable benefit without paying the cost of exhaustive full-program search. Moreover, the overlap ratio indicates that the Score Agent does not simply reproduce the PGO hot-function set. Instead, it re-ranks optimization candidates according to their actual contribution to end-to-end speedup. 
\begin{tcolorbox}[colback=olive!20,colframe=black,arc=3pt,boxrule=0.8pt]
\textbf{\textit{RQ4:}} \textit{On benchmarks that exceed the LLM context limit, the Score Agent consistently matches or outperforms PGO-based function selection by re-ranking functions according to their actual contribution to end-to-end speedup rather than profile-derived hotness alone.}
\end{tcolorbox}

\begin{table}[t]
\centering
\footnotesize
\caption{Agent ablation analysis. }\label{tab:ablation}
\vspace{-3mm}
\begin{tabular}{lccc}
\toprule
\textbf{Configuration} & \textbf{Round 1} & \textbf{Round 2} & \textbf{Round 3} \\
\midrule
No Evaluation Agent & - & 0.961$\pm$0.181 & 0.961$\pm$0.122 \\
No Reasoning Agent & 0.910$\pm$0.135 & 0.823$\pm$0.360 & 0.870$\pm$0.301 \\
No Analysis Agent   & 0.969$\pm$0.113 & 0.977$\pm$0.135 & 1.019$\pm$0.089 \\
\textbf{\SystemName (Full)} & \textbf{1.010$\pm$0.121} & \textbf{1.016$\pm$0.125} & \textbf{1.020$\pm$0.116} \\
\bottomrule
\end{tabular}%
\vspace{-4mm}
\end{table}

	 \vspace{-3mm}
\subsection{Ablation Study (RQ5)} 
We study component contributions by disabling the Analysis, Reasoning, and Evaluation Agents in turn (Table~\ref{tab:ablation}). The Reasoning Agent is the most critical for optimization effectiveness: without it, performance starts at only 0.910$\times$ in Round~1 and drops further to 0.823$\times$ in Round~2, indicating that pass selection without explicit reasoning can significantly hurt performance. The Evaluation Agent mainly supports robustness. Without it, the first round remains the same as the full \SystemName system, but performance declines from 1.010$\times$ to 0.961$\times$ in later rounds, showing that iterative refinement becomes unstable without corrective evaluation. The Analysis Agent primarily improves convergence efficiency. When it is removed, the system starts from a weaker state (0.969$\times$) and spends early rounds recovering from less informed choices before approaching near-optimal performance by Round~3.
\begin{tcolorbox}[colback=olive!20,colframe=black,arc=3pt,boxrule=0.8pt]
\textbf{\textit{RQ5:}} \textit{The Reasoning Agent is most critical for optimization effectiveness, the Evaluation Agent for robustness, and the Analysis Agent for convergence efficiency.}
\end{tcolorbox}

\subsection{Trace-Driven Case Study (RQ6)}
\label{sec:rq2_case_study}
To show that \SystemName makes evidence-based and interpretable decisions, we present representative internal inference traces for Qsort on the x86-64 platform. Since the Qsort input fits within the LLM context limit, this case does not require Score Agent analysis.

\cparagraph{Step 1: Analysis Agent --- grounded diagnosis from IR and compiler evidence.}
The Analysis Agent is configured to read the actual LLVM IR and produce a concise natural-language summary of the program structure and likely bottlenecks. In the original trace, the prompt asks the agent to generate an \texttt{ir\_analysis\_summary} from the IR content. 
To keep the trace compact, we preserve only the key task intent:

\begin{lstlisting}[basicstyle=\ttfamily\scriptsize, breaklines=true, frame=single, caption={Analysis Agent Prompt (abridged)}]
[Round 1] Generate a concise natural-language analysis summary from the provided LLVM IR. Focus on program structure and optimization-relevant bottlenecks.
\end{lstlisting}

The resulting analysis identifies the workload as memory-intensive and loop-heavy, with the sorting routine as the main optimization target. This semantic interpretation is then paired with compiler evidence extracted by tools. The actual feature payload shows why the agent focuses on \texttt{qsortx}: it has \textbf{198 blocks}, \textbf{12 calls}, and \textbf{16 loops}. More importantly, the compiler remarks reveal repeated missed vectorization opportunities, including \textbf{13} \texttt{slp-vectorizer} failures marked \texttt{NotBeneficial},
which directly informed the decision to adjust the `slp-threshold'.

\begin{lstlisting}[basicstyle=\ttfamily\scriptsize, breaklines=true, frame=single, caption={Evidence Generated by Analysis Agent (abridged)}]
  "name": "qsortx",
  "features": {
    "callee": { "Blocks": 198, "Calls": 12 },
    "generic": { "num_loops": 16 }
  },
  "remarks": {
    "missed": {
      "slp-vectorizer": [
        { "count": 13, "Name": "NotBeneficial" }
      ],
      "loop-vectorize": [
        { "count": 3, "Name": "NonReductionValueUsed" }
      ]
    },
    "passed": {
      "loop-vectorize": [
        { "count": 4, "Name": "Vectorized" }
\end{lstlisting}

\paragraph{Step 2: Reasoning Agent --- an evidence-grounded optimization decision.}
The Analysis Agent passes the natural-language summary, the extracted feature, and the remark payload to the Reasoning Agent. It then provides an LLVM optimization strategy from these inputs and the target platform information. The original trace shows that the Reasoning Agent receives the IR diagnosis, static features, compiler remarks, and the x86-64 platform description before generating its optimization JSON. 
\vspace{-2mm}
\begin{lstlisting}[basicstyle=\ttfamily\scriptsize, breaklines=true, frame=single, caption={Reasoning Agent Prompt (abridged)}]
Input:
- baseline -O3 pass pipeline
- IR analysis summary
- static features and compiler remarks
- previous-iteration runtime and counter data
- target platform: x86-64 / Intel i9-11900K

Task:
Generate an optimized LLVM pass pipeline and parameter settings.
Explain the rationale using the observed bottlenecks.
\end{lstlisting}

The actual Round-1 output is aggressive. It raises \texttt{unroll\_count} to 8, \texttt{unroll\_threshold} to 600, \texttt{inline\_threshold} to 800, and lowers \texttt{slp\_threshold} to \(-5\) (lowering it makes the vectorizer more willing to apply SLP vectorization). The rationale states that these changes aim to prioritize loop and memory optimizations, increase loop unrolling for the sorting kernel, and make SLP vectorization more aggressive in response to the observed missed opportunities. 
\vspace{-2mm}
\begin{lstlisting}[basicstyle=\ttfamily\scriptsize, breaklines=true, frame=single, caption={Actual Reasoning Agent Output (Round 1, abridged)}]
  "passes_param_adjustments": {
    "unroll_count": 8,
    "unroll_threshold": 600,
    "inline_threshold": 800,
    "slp_threshold": -5,
    "mcpu": "skylake",
    "mtriple": "x86-64-unknown-linux-gnu"
  },
  "confidence": 0.9,
  "rationale": "Increase loop unrolling and make SLP vectorization
  more aggressive to address loop bottlenecks and missed SLP
  opportunities in qsortx."
\end{lstlisting}

\paragraph{Step 3: Evaluation Agent --- measured evidence exposes why the decision was harmful.}
The Evaluation Agent compiles the new pipeline, runs performance measurement, and is explicitly constrained to use \emph{real} runtime data.

\begin{lstlisting}[basicstyle=\ttfamily\scriptsize, breaklines=true, frame=single, caption={Evaluation Agent Prompt (abridged)}]
[Round 1] Analyze the optimization result using the received
runtime data.

Critical instruction:
- Use the real data only.
- Compare the new version against the baseline.
- Identify the cause of speedup or slowdown.
- Find the next-round optimization points.
\end{lstlisting}

The measured Round-1 result shows that the aggressive policy is harmful. Execution time increases from \textbf{1.5355} to \textbf{1.5474s} (\(+0.77\%\)). L1 misses jump from about \textbf{0.96M} to \textbf{2.25M} (\(+133.1\%\)), and IPC drops. The Evaluation Agent attributes this regression to excessive unrolling and overly permissive vectorization, concluding that the policy exhibits \emph{misaligned aggressiveness} on x86-64. 

\begin{lstlisting}[basicstyle=\ttfamily\scriptsize, breaklines=true, frame=single, caption={Actual Evaluation Agent Diagnosis (Round 1, abridged)}]
{
  "analysis_report": {
    "analysis": "The Round 1 optimization results show a performance regression of 0.77% in average time elapsed (1.5355s to 1.5474s), despite a 2.3% reduction in total instruction count (28.69B to 28.03B). This indicates a significant drop in execution efficiency... The optimization is too aggressive: unroll-count=8 and unroll-threshold=600 cause excessive code expansion , while slp-threshold=-5 is too permissive ."
  }
}
\end{lstlisting}

\paragraph{Step 4: Feedback generation --- turning diagnosis into a corrective next action.}
The Evaluation Agent then compresses its diagnosis into a concise instruction for the next round. The original trace explicitly asks for a brief summary containing the optimization result, the likely bottleneck, and the next-round adjustment directions, while forbidding the introduction of unrelated new passes or parameters. 
\vspace{-4mm}
\begin{lstlisting}[basicstyle=\ttfamily\scriptsize, breaklines=true, frame=single, caption={Feedback Prompt (abridged)}]
Based on the evaluation report, generate a brief optimization
summary for the next round:
- what happened and why,
- the likely bottleneck,
- which parameters should be more conservative or more aggressive.
\end{lstlisting}

\begin{tcolorbox}[colback=olive!20,colframe=black,arc=3pt,boxrule=0.8pt]
\textbf{\textit{RQ6:}} \textit{The trace-driven case study shows that grounding makes \SystemName's optimization decisions interpretable and diagnosable by linking pipeline edits, regressions, and corrections to concrete compiler evidence and runtime feedback.}
\end{tcolorbox}

The feedback guides the next round toward a more conservative policy by reducing unrolling aggressiveness and avoiding non-beneficial vectorization. This is the critical transition from \emph{diagnosis} to \emph{repair}: the system does not merely reject the previous round, but uses measured evidence to produce a concrete correction path. The trace shows that this recovery succeeds: Round 2 largely removes the regression, and by Round 3 the execution time improves to 1.4941s, corresponding to a \textbf{1.028$\times$} speedup over the \texttt{-O3} baseline.

\section{Performance of Reasoning Backend Generalizability\label{sec:general}}
We also evaluate the \SystemName across four distinct LLMs as the backend: DeepSeek-V3.2, ChatGPT-4o~\cite{openai2024gpt4o}, Qwen3~\cite{yang2025qwen3}, and Gemini 3 Flash~\cite{google2025gemini3flash}. Table~\ref{tab:performance_wo_rb_appendix} reports cBench results on x86-64 and ARM64 without the rollback policy. The data proves that \SystemName is a robust, model-agnostic optimization framework, with all evaluated LLMs achieving runtime performance improvements in R3. Specifically, DeepSeek-V3.2 shows a strong initial reasoning (R1 Geo. Mean: 1.010$\times$ on x86), Gemini 3 Flash on ARM64 initially exhibits a significant regression (0.922$\times$, with 21 losses). However, by Round 3, the feedback loop successfully corrects these errors, guiding the system to a 1.091$\times$ speedup.  This process proves that \SystemName's iterative correction mechanism effectively compensates for variance in LLM reasoning baselines.

\newcommand{\nogray}{\cellcolor{white}}

\begin{table}[t]
    \centering
        \caption{Performance comparison without rollback policy on cBench with DeepSeek-V3.2, ChatGPT 4o, Qwen3 and Gemini 3 Flash as backend. The $\pm$ denotes the standard deviation of speedups across the benchmarks in the cBench suite.} 
        \vspace{-3mm}
    \label{tab:performance_wo_rb_appendix}
    
    \footnotesize
    \begin{tabular}{l l l r}
        \toprule
        \textbf{Platform} & \textbf{Model} & \textbf{Method} & \textbf{Geo. Mean} \\
        \midrule
        \rowcolor{gray!15}\nogray 

        & & \SystemName (best in R3)& 1.040$\pm$0.114\\
       \rowcolor{gray!15}\nogray 
        & \multirow{-2}{*}{DeepSeek-V3.2} & \SystemName (R1) & 1.010$\pm$0.121 \\
        
        & & \SystemName (best in R3)& 1.029$\pm$0.105 \\
        & \multirow{-2}{*}{ChatGPT 4o} & \SystemName (R1) & 0.992$\pm$0.128 \\

           \rowcolor{gray!15}\nogray 
        & & \SystemName (best in R3)& \cellcolor{bestgreen}{\textbf{1.040$\pm$0.105}} \\

           \rowcolor{gray!15}\nogray 
        &\multirow{-2}{*}{Qwen3} & \SystemName (R1) & 0.995$\pm$0.116 \\
        
        & & \SystemName (best in R3)& \second{1.040$\pm$0.099} \\
        
        \multirow{-8}{*}{\textbf{x86-64}} & \multirow{-2}{*}{Gemini 3 Flash} & \SystemName (R1) & 1.008$\pm$0.105 \\

        \midrule
           \rowcolor{gray!15}\nogray  
        & & \SystemName (best in R3)&\cellcolor{bestgreen}{\textbf{ 1.109$\pm$0.206}}\\
           \rowcolor{gray!15}\nogray 
        & \multirow{-2}{*}{DeepSeek-V3.2} & \SystemName (R1) & 1.004$\pm$0.215 \\  

        & \multirow{2}{*}{ChatGPT 4o} & \SystemName (best in R3)& 1.088$\pm$0.211 \\
        & & \SystemName (R1) & 0.989$\pm$0.242 \\

           \rowcolor{gray!15}\nogray 
        &  & \SystemName (best in R3)& 1.080$\pm$0.227 \\
           \rowcolor{gray!15}\nogray 
        & \multirow{-2}{*}{Qwen3} & \SystemName (R1) & 0.975$\pm$0.193 \\
        
        & & \SystemName (best in R3)& \second{1.091$\pm$0.225} \\
        \multirow{-8}{*}{\textbf{ARM64}} & \multirow{-2}{*}{Gemini 3 Flash} & \SystemName (R1) & 0.922$\pm$0.199 \\
        
        \bottomrule
    \end{tabular}

    \vspace{-5mm}
\end{table}

\vspace{-2mm}
\section{Related Work}
\vspace{-2mm}
\cparagraph{Classical and Iterative Compilation.}
Modern compilers such as LLVM~\cite{lattner2004llvm} and GCC~\cite{GCCInternals} rely on fixed, expert-crafted optimization pipelines (e.g., -O3) to manage the complex optimization configuration problem. While effective for general use, these static optimization policies often miss program-specific opportunities. To address this, iterative compilation frameworks like OpenTuner, TVM~\cite{chen2018tvm}, and genetic algorithms have been developed to search for optimal configurations. However, these methods treat the compiler as a black box and require thousands of computationally expensive recompilations. Furthermore, the resulting pipelines are often overfitted to specific benchmarks or hardware, lacking the semantic insight required to generalize effectively to new code without extensive retraining.

\cparagraph{Profile-Guided and ML-Based Optimization.} 
Industry-standard approaches, including Instrumentation-based PGO and AutoFDO, improve upon static heuristics by utilizing runtime execution profiles to guide decisions such as inlining and block placement. However, these methods remain constrained within the fixed topology of the default pipeline and are sensitive to profile quality. The machine learning approaches like MLGO~\cite{trofin2021mlgo} and ACPO~\cite{ashouri2023acpo} have demonstrated success in learning specific heuristics. Yet, ML-based models typically target narrow decision spaces or single passes, leaving the potential gains from holistic, whole-program pipeline reordering largely unexplored. 
Specifically, Reinforcement Learning frameworks like Autophase~\cite{huang2019autophase}  and CompilerGym~\cite{cummins2022compilergym} have advanced the state of the art in phase ordering. However, they fundamentally rely on heavy offline training phases, often consuming weeks of GPU time to learn a generalized policy from millions of compilation traces. These methods struggle to adapt to unseen workloads without extensive retraining. In contrast, \SystemName functions as a zero-shot, inference-only system. It eliminates the training overhead entirely by treating optimization not as a pattern-matching task, but as a reasoning task.

\cparagraph{LLMs for Code Optimization.}
The emergence of LLMs has introduced semantic reasoning into program optimization, complementing traditional search- and heuristic-based methods. Recent studies use LLMs to generate optimizations directly~\cite{hu2025compileagent,cummins2024meta}, incorporate stronger correctness guarantees through verification-guided learning and validated transformations~\cite{fang2026llm,taneja2025llm}, or perform source-level transformations for specific optimization tasks such as vectorization and parallelization~\cite{zheng2025vectrans}. Beyond optimization, a growing body of work shows that LLMs are effective for related code tasks, including decompilation~\cite{she2024wadec}, iterative code generation~\cite{fu2025first}, SIMD-oriented code synthesis~\cite{he2025simdbench}, autonomous program repair~\cite{bouzenia2025repairagent}, and search-based code optimization~\cite{gao2024search}. Recent multi-agent frameworks further suggest that role specialization and iterative self-reflection can improve performance on complex programming tasks such as code generation and automated testing~\cite{liu2025temac,pan2501codecor}. Nevertheless, these methods largely operate at the source-code level or target isolated code-generation tasks, and therefore do not directly address compiler pass-pipeline optimization under strict compiler validity constraints and architecture-dependent performance objectives. In contrast, \SystemName targets inference-only LLVM pass-pipeline optimization for runtime performance, performing constrained pass edits inside the compiler loop and refining them iteratively with compiler evidence and measured execution feedback.

\vspace{-3mm}
\section{Conclusion}
We present \SystemName, an inference-only multi-agent framework for LLVM compiler performance tuning. \SystemName operates in an inference-only manner: it performs constrained pass-pipeline edits, validates them deterministically, and iteratively refines them using measured execution behavior. In this way, it occupies a practical middle ground between conservative compiler heuristics and expensive autotuning. Experiments show that \SystemName outperforms industrial PGO baselines and budget-constrained OpenTuner, and the compiler evidence makes optimization decisions more interpretable and diagnosable. More broadly, this work points to a promising direction for software engineering: combining existing compiler infrastructure with LLM-based reasoning to build practical, adaptive, and explainable optimization workflows.
\vspace{-2mm}

\bibliographystyle{named}
\bibliography{refs}

\end{document}